\documentclass[final,authoryear,5p,times,twocolumn]{elsarticle}

\usepackage{graphicx}
\usepackage{amssymb,amsmath}

\journal{Journal of Theoretical Biology}

\begin{document}

\begin{frontmatter}

\title{Stochastic amplification in an epidemic model with seasonal forcing}

\author{Andrew J. Black\corref{cr1}}
\ead{andrew.black@postgrad.manchester.ac.uk}
\cortext[cr1]{Phone: +441612754201}

\author{Alan J. McKane}

\address{Theoretical Physics Group, School of Physics and Astronomy, 
University of Manchester, Manchester, M13 9PL, UK}

\begin{abstract}
We study the stochastic susceptible-infected-recovered (SIR) model with
time-dependent forcing using analytic techniques which allow us to disentangle 
the interaction of stochasticity and external forcing. The model is
formulated as a continuous time Markov process, which is decomposed into a 
deterministic dynamics together with stochastic corrections, by using an 
expansion in inverse system size. The forcing induces a limit cycle in the 
deterministic dynamics, and a complete analysis of the fluctuations about 
this time-dependent solution is given. This analysis is applied when the 
limit cycle is annual, and after a period-doubling when it is biennial. The 
comprehensive nature of our approach allows us to give a coherent picture 
of the dynamics which unifies past work, but which also provides a 
systematic method for predicting the periods of oscillations seen in 
whooping cough and measles epidemics.
\end{abstract}

\begin{keyword}
non-linear dynamics \sep period doubling \sep measles
\end{keyword}

\end{frontmatter}

\section{Introduction}
The availability of extensive time-series data for childhood diseases is often 
the reason given for the amount of attention that this subject receives.
However, the possibility that relatively simple models can capture the 
essence of the disease dynamics also makes the topic an attractive one for 
modellers. Mathematical epidemiologists are especially intrigued by the rich 
variety of oscillatory dynamics seen in this data \citep{earn_science,GBF02}. 
Within the literature there is a broad consensus that there are two main 
elements needed to model these oscillations: firstly stochasticity, due to 
the individual nature of the population \citep{BMS60,DL94}; and secondly, 
seasonal forcing, arising from the term-time aggregation of children in 
schools, which is deterministic \citep{LY73,Sch85,ADHP+06}. Independently 
these two factors are well understood, but how they interact when both 
included in the same model is still an open question \citep{KRG01,RKG02,CRP04}.

Measles is the canonical example of a disease which displays recurrent 
epidemic behaviour. In larger cities regular periodic oscillations, usually 
annual or biennial, are observed, whereas smaller cities display more 
irregular dynamics \citep{GBF02,LS}. The introduction of mass vaccination in 
the 1960s provides a `natural experiment' after which the dynamics become
much more irregular \citep{GH97}. One of the early successes in the field was 
a simple deterministic model with seasonal forcing which could recreate the 
regular dynamics of measles \citep{Die76,SS83}. Where simple models such 
as this fail, is in capturing the more irregular dynamics seen in smaller 
populations \citep{GBF02}, after vaccination \citep{REG99}, and in other 
diseases such as whooping cough \citep{NR08}. These aspects can only be 
captured by fully stochastic models. 

Stochastic models without external forcing show large oscillations caused by 
the stochasticity exciting the system's natural 
frequency \citep{bartlett2,alonso}. When forcing is included, it is less 
clear how the stochasticity interacts with the cyclic solutions that are 
produced. It could act passively to kick the system between different 
deterministic states \citep{Sch85b}, as well as interacting with the 
non-linearity to excite the transients. Power spectra \citep{Pri82,AGM84}
have proved a useful tool in investigating these factors. They can help 
distinguish various components in the time-series and classify them as 
essentially seasonal, stochastic or an interaction of the two \citep{Ben06}. 
The most successful synthesis, by \citet{BE03}, showed that a simple 
mechanistic model can accurately predict the position of peaks in the power 
spectrum of a number of different disease time series. 

We approach this problem by starting with an individual based model, which is 
inherently stochastic. We can then both simulate it and derive the emergent 
population level dynamics. The novel aspect of this work is that we calculate 
the power spectrum analytically for explicitly time-dependent external forcing,
and compare the results with stochastic simulations. We do this by formulating 
the model as a master equation which can then be studied using van 
Kampen's \citeyearpar{van_kampen} expansion in the inverse system size. The 
macroscopic dynamics can then be viewed as a sum of a deterministic and a 
stochastic part. The value of the analytic approach is that we can more easily 
deduce the mechanisms behind the dynamics and better understand the interplay 
between deterministic and stochastic forces.

The theory we develop in this paper unifies much of the previous work on 
these models. It encompasses the influential work of \citet{earn_science} in 
understanding the transitions in measles epidemics, the later work 
of \citet{BE03} relating to the transient fluctuations close to cyclic 
attractors for different diseases and the more recent work on stochastic 
amplification in epidemic models \citep{alonso, BM10}. The picture that emerges
is close to that proposed by \cite{BE03}, but goes beyond it in two important 
respects. Firstly, we calculate the exact power spectrum for the forced model. 
Secondly, we show how the forcing changes the form of the fluctuations, and
how in a stochastic model these are intimately related to the period doubling 
bifurcation, which is vital for explaining the dynamics of measles.

The rest of this paper is as follows. In section \ref{sec:model} we introduce 
the seasonally-forced version of the stochastic susceptible-infected-recovered 
(SIR) model and the system-size expansion of the master equation. 
Section \ref{sec:results} provides a discussion of the results for the simple 
case when the deterministic dynamics are described by an annual limit cycle. 
In Section \ref{sec:measles} we apply our method to elucidate the dynamics 
investigated by \cite{earn_science}, which can account for the transitions 
in measles epidemics. This is an interesting parameter regime, as the 
deterministic theory predicts a period doubling bifurcation. Finally, in
section \ref{sec:discussion} a broad discussion of our results is given, 
describing how this approach can account for the different dynamics of measles 
and whooping cough. There are two appendices giving technical details relating
to the system-size expansion and Floquet theory.

%%%%%%%%%%%%%%%%%%%%%%%%%%%%%%%%%%%%%%%%%%%%%%%%%%%%%%%%%%%%%%%%%%%%%%%%%%%%%%
%%%%%%%%%%%%%%%%%%%%%%%%%%%%%%%%%%%%%%%%%%%%%%%%%%%%%%%%%%%%%%%%%%%%%%%%%%%%%%

\section{The seasonally forced SIR model}
\label{sec:model}

We first summarise the individual-based stochastic SIR model. We emphasise
only the aspects which are relevant to this paper; a more general discussion 
can be found in textbooks on the subject \citep{and_may,KR07}. The population 
is split into three classes: susceptibles, infected and recovered. Birth and 
death rates are set equal to $\mu$ and these events are linked, even in the 
stochastic model, so that the total population, $N$, remains constant. Recovery
happens at a constant rate $\gamma$, so that the average infectious time is 
$1/\gamma$; once recovered, an individual is immune for life. Seasonal forcing 
is included by assuming that the transmission rate $\beta(t)$ follows a 
term-time pattern \citep{Sch85},
\begin{equation}
\beta(t)=\beta_0 (1+\beta_1 \text{term}(t)),
\end{equation}
where $\beta_0$ is the baseline contact rate, $\beta_1$ the magnitude of 
forcing and $\text{term}(t)$ is a periodic function which switches between $1$ 
during school terms and $-1$ during holidays. In this paper we use the England 
and Wales term dates set down by \cite{KRG01}. The reproductive ratio is
determined by $R_0=\langle \beta \rangle / \gamma$, where 
$\langle \beta \rangle$ is the effective (time-averaged) transmission rate:
\begin{equation}
\langle \beta \rangle = \beta_0 [ p_s (1+\beta_1) +(1-p_s)(1-\beta_1)],
\end{equation}
and $p_s$ is the proportion of time spent in school. For our choice of terms 
$p_s=0.75$. We also include a small immigration term, $\eta$, to account for 
infectious imports. We use a commuter formulation, where susceptibles are in 
contact with a pool of infectives outside the main 
population \citep{ED94,alonso}. Since $N=S+I+R$, we can use this constraint 
to eliminate the variable $R$ from the rate equations.

The model is then defined by the processes through which it evolves. If we 
write the state of the system as $\sigma\equiv\{S,I\}$, we can specify the 
following transition rates, $T(\sigma | \sigma')$, between an initial state 
$\sigma'$ and a final state $\sigma$:
\begin{enumerate}
\item Infection: $S+I \xrightarrow{\beta(t)}I+I$ and $S \xrightarrow{\eta}I$.
\begin{equation}T(S-1,I+1|S,I)=\Bigl(\beta(t)\frac{S}{N}I + \eta S\Bigr).
\end{equation}
\item Recovery: $I\xrightarrow{\gamma}R$.
\begin{equation}T(S,I-1|S,I)=\gamma I. 
\end{equation}
\item Death of an infected individual: $I \xrightarrow{\mu}S$.
\begin{equation}T(S+1,I-1|S,I)=\mu I. \end{equation}
\item Death of a recovered individual: $R \xrightarrow{\mu}S$.
\begin{equation}T(S+1,I|S,I)=\mu(N-S-I). \end{equation}
\end{enumerate}
Since birth and death are coupled, the processes 3 and 4 also imply the birth
of a susceptible individual. An important point is that changes in vaccination 
can be mapped onto the effective transmission rate $\langle\beta\rangle$ 
using \citep{earn_science}
\begin{equation}
\langle\beta\rangle\rightarrow \langle\beta\rangle (1-p),
\end{equation}
where $p$ is the proportion of individuals vaccinated at birth. We can also 
approximately map a change in birth rates onto $\langle\beta\rangle$, but this 
is not exact in this model because births and deaths are linked. In this paper 
we are primarily interested in the parameter range of childhood diseases. 
These are characterised by $\mu \ll \gamma$, \textit{i.e.}~the average life 
expectancy of an individual is orders of magnitude longer than the mean 
infectious period \citep{and_may}.

\subsection{Methods of analysis}

We use two methods to investigate the dynamics of this system. Firstly we 
simulate the system using Gillespie's \citeyearpar{gillespie} algorithm with 
the appropriate time-dependent extensions \citep{And07}. This method generates 
exact realisations from which statistical quantities, such as power spectra
and moments, can be computed. The second method is analytic, through the 
construction of a master equation. The master equation describes the evolution 
of the probability distribution of finding the system in state $\sigma$ at 
time $t$,
\begin{equation}
\frac{dP(\sigma;t)}{dt} = \sum_{\sigma' \neq \sigma} T(\sigma | \sigma')
P(\sigma';t) - \sum_{\sigma' \neq \sigma} T(\sigma' | \sigma) P(\sigma;t).
\label{master}
\end{equation}
This cannot be solved exactly so we instead use 
van Kampen's \citeyearpar{van_kampen} expansion in the inverse system size to 
derive approximate analytic solutions. This involves making the substitutions,
\begin{equation}
\begin{aligned}
S=N\phi+N^{1/2}x,\\
I=N\psi+N^{1/2}y,
\end{aligned}
\label{ansatz}
\end{equation}
and expanding the master equation in powers of $N^{-1/2}$. This technique and 
similar ones have been documented at length in the literature, but almost 
exclusively for time-independent models \citep{alonso}. The novel aspect of 
this paper is that we analyse the full time-dependent system \citep{BGM09b}, 
whereas in a previous paper we used the approximation which replaced $\beta(t)$
by $\langle\beta\rangle$ \citep{BM10}.

The details of the system-size expansion for this model are given in Appendix 
A. At leading order we find a pair of deterministic equations, describing the 
mean behaviour, which scale with the system size $N$, 
\begin{equation}
\label{eq:macro}
\begin{aligned}
\dot{\phi}= -\beta(t)\phi\psi -\eta\phi+\mu(1-\phi), \\
\dot{\psi}= \beta(t)\phi\psi +\eta\phi-(\mu+\gamma)\psi.
\end{aligned}
\end{equation}
These are the same as the equations that are found from a purely 
phenomenological treatment of the SIR system. At next-to-leading order we 
obtain a pair of Langevin equations for the stochastic corrections to the 
deterministic equations \eqref{eq:macro},
\begin{equation}
\label{eq:langevin}
\dot{\mathbf x}=K(t)\mathbf{x}(t)+\mathbf{f}(t),
\end{equation}
where $\mathbf{x}\equiv\{x,y\}$, and $\mathbf{f}(t)$ are Gaussian white-noise 
terms with correlation function 
$\langle \mathbf{f}(t)\mathbf{f}(t')^{T} \rangle=G(t)\delta(t-t')$. The 
matrices $K(t)$ and $G(t)$ are determined from carrying out the expansion 
and are given by
\begin{equation}
K(t)=\begin{pmatrix}
-\beta\bar{\psi}-\eta-\mu & -\beta\bar{\phi}\\
\beta\bar{\psi}+\eta & \beta\bar{\phi}-\gamma-\mu
\end{pmatrix},
\label{K_matrix}
\end{equation}
and
\begin{equation}
\begin{aligned}
G_{11}&=\beta\bar{\phi}\bar{\psi}+\eta\bar{\phi}+\mu(1-\bar{\phi}),\\
G_{22}&=\beta\bar{\phi}\bar{\psi}+\eta\bar{\phi}+(\gamma+\mu)\bar{\psi},\\
G_{12}=G_{21}&=-\beta\bar{\phi}\bar{\psi}-\eta\bar{\phi}-\mu\bar{\psi},\\
\end{aligned}
\label{G_matrix}
\end{equation}
where a bar indicates that the solutions are evaluated on the limit cycle. 
These are essentially the same as are found in the non-forced 
model \citep{alonso}, except now $\beta$, $\bar{\phi}$ and $\bar{\psi}$ are 
all functions of time.

%%%%%%%%%%%%%%%%%%%%%%%%%%%%%%%%%%%%%%%%%%%%%%%%%%%%%%%%%%%%%%%%%%%%%%%%%%%%%
%%%%%%%%%%%%%%%%%%%%%%%%%%%%%%%%%%%%%%%%%%%%%%%%%%%%%%%%%%%%%%%%%%%%%%%%%%%%%

\section{Stochastic amplification about a limit cycle}
\label{sec:results}

\begin{figure}[ht]
\centering
\includegraphics[width=\columnwidth]{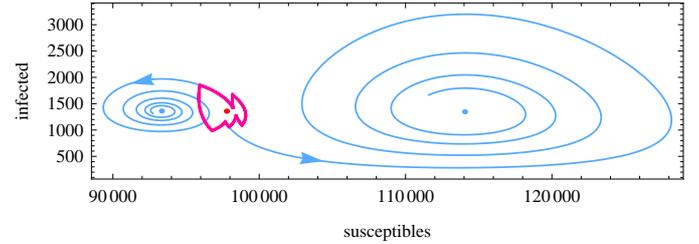}
\caption{Phase portrait illustrating a deterministic solution of the forced 
SIR model. The term-time forcing creates a limit cycle (red curve) as the 
system alternately spirals between the two fixed points defined by 
$\beta_H=\beta_0(1+\beta_1)$ and $\beta_L=\beta_0(1-\beta_1)$. The light blue 
solutions show the behaviour if the forcing was switched off, to illustrate 
the two spiral attractors.  The red dot shows the fixed point calculated using 
the approximation where $\beta(t)$ is replaced by $\langle\beta\rangle$.}
\label{fig:new_att}
\end{figure}

The mean behaviour is found by integrating the deterministic equations
\eqref{eq:macro}. When $\beta_1=0$, solutions show damped oscillations tending 
to a fixed point \citep{and_may}. For non-zero $\beta_1$, this model can 
display a rich set of dynamics including chaos \citep{OTS88,RW91}, but for 
realistic parameter values the most common long-time solution is a 
limit cycle with a period that is an integer multiple, $n$, of a 
year \citep{Die76,SS83}. As the forcing is a step function in time, 
we can visualise this as the system alternately switching between two spiral 
fixed-points \citep{KRG01} resulting in a piecewise continuous limit cycle, 
illustrated in figure \ref{fig:new_att}. Any other periodic forcing function,
for instance a sinusoidally varying one, could be used without more difficulty,
and would typically lead to a limit cycle which is smooth. As $\beta_1$ is 
increased, the limit cycle grows (although typically not linearly with 
$\beta_1$) and at critical values bifurcations are induced to longer period 
solutions \citep{AS84,KP94}.

In this section we present results where there is only an annual limit-cycle 
($n=1$). The case where we also have a period doubling is examined in 
section \ref{sec:measles}. The stability of these limit-cycle solutions can 
be investigated with the use of Floquet theory \citep{Kuz04,BGM09b}. This 
quantifies how perturbations to the trajectory of the limit cycle behave and 
is analogous to linear stability analysis about a fixed point \citep{Gri90}.

Floquet theory states that for any periodic solution of Eq.~\eqref{eq:macro}
there exists a matrix $B$ which satisfies the relation,
\begin{equation}
X(t+T_n)=X(t)B,
\end{equation}
where $X(t)$ is the fundamental matrix \citep{Gri90} and $T_n$ is the period 
of the limit cycle. The eigenvalues of $B$ are called the Floquet multipliers, 
$\rho_i$; a related set of quantities are the Floquet exponents 
$\lambda_i=\ln(\rho_i)/T_n$ (in this paper, since we will be discussing 
frequencies rather than angular frequencies, these exponents will be divided 
by a factor of $2\pi$). Another way to think of this is as linear stability 
analysis of the fixed points of the $n$-cycle Poincare map of the 
system \citep{BE03,Kuz04}. A limit-cycle solution will be stable if 
$|\rho_i|<1$. When the multipliers are complex, perturbations to the 
trajectories return to the limit-cycle in a damped oscillatory manner, 
analogous to a stable spiral fixed point \citep{Gri90}. Similar ideas have 
been used to investigate the transients in forced epidemic systems in the 
past, but only in a deterministic setting \citep{BE03,HDE07}. Here we will 
explore how the nature of the fluctuations can be quantified using Floquet 
theory.

\begin{figure}[ht]
\centering
\includegraphics[width=\columnwidth]{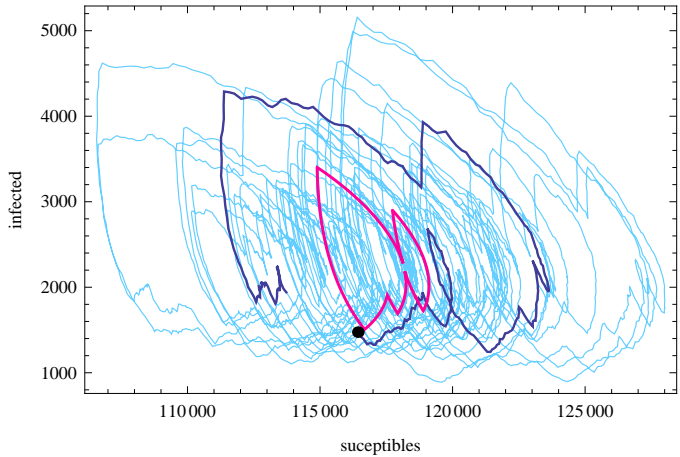}
\caption{Phase portrait of the stochastic SIR system. A time-series of 100 
years duration is shown in light blue. The first two years are highlighted in 
dark blue, 
with the dot showing the start point. The macroscopic limit-cycle (red) is 
also superimposed. Parameters are those relevant for whooping cough 
 \citep{NR08}: $R_0=17$, $\gamma=1/22$, $\beta_1=0.25$, 
$\mu=5.5\times 10^{-5}$, $\eta=10^{-6}$ and $N=2\times 10^6$.}
\label{fig:stoch_ex}
\end{figure}

Figure \ref{fig:stoch_ex} shows a simulation of the full stochastic system 
together with the deterministic limit-cycle solution. We can see that even for 
large populations the stochastic corrections to the deterministic solution are 
important. The noise due to demographic stochasticity (noise at the individual 
level due to chance events; \citealt{nisbet}) excites the natural oscillatory 
modes about the limit cycle, creating a resonance and giving rise to large 
scale coherent oscillations---an effect known as stochastic 
amplification \citep{mckane_new2,alonso}. As described in Appendix B, by 
solving Eq.~\eqref{eq:langevin}, and invoking aspects of Floquet theory, we 
can express the auto-correlation function,
\begin{equation}
C(\tau)=\frac{1}{T_n}\int_0^{T_n} 
\langle \mathbf{x}(t+\tau)\mathbf{x}^{T}(t)\rangle \,dt, \ \
\mathbf{x}\equiv\{x,y\},
\end{equation}
as an integral without further approximation \citep{BGM09b}. Taking the 
Fourier transform of this expression then gives an exact expression for power 
spectrum of these stochastic oscillations.

\begin{figure}[ht]
\centering
\includegraphics[width=\columnwidth]{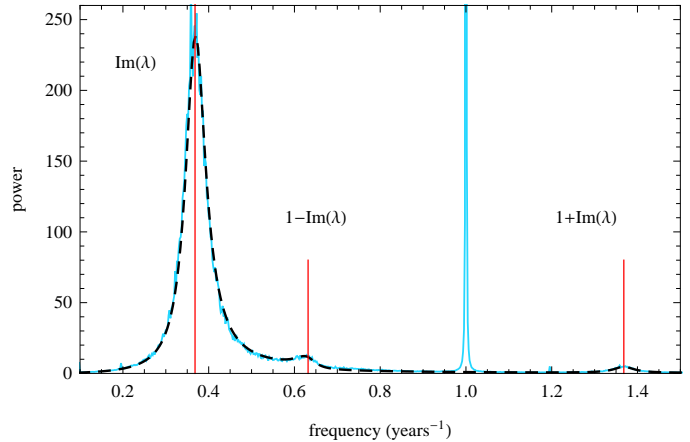}
\caption{Power spectra for the number of infectives from simulation (light 
blue solid curve) and analytic calculation (black dashed curve). From the 
simulations, we observe a sharp peak at 1 year from the deterministic 
annual-limit cycle. The other peaks, marked by the red lines, are from 
stochastic amplification, with the peak frequencies given by 
$m\pm\text{Im}(\lambda)$, where $\text{Im}(\lambda)=0.36$. The dominant 
stochastic period is therefore $1/0.36=2.7$ years. Parameters are as in 
figure \ref{fig:stoch_ex}}
\label{fig:example}
\end{figure}

Figure \ref{fig:example} shows simulated and analytic power spectra for the 
system shown in figure \ref{fig:stoch_ex}. We observe a sharp peak at 1 year 
due to the deterministic annual limit-cycle and a number of  broader peaks due
to the stochastic amplification of the transients. We would expect on general
grounds that the stochastic peaks would be observed at frequencies,
\begin{equation}
m /T_n\pm {\rm Im}(\lambda),
\end{equation}
where $m$ is an integer and $\lambda$ is the Floquet 
exponent \citep{Wis85,BGM09b}, and this is indeed what is seen. For the annual 
limit-cycle the dominant peak is at $0+\text{Im}(\lambda)$, with the other 
peaks being much smaller. Near to bifurcations these minor peaks become 
important and are treated in more detail in the following section.

The area under the peaks in the power spectrum is proportional to the 
root-mean-square amplitude of the oscillations. Away from any deterministic 
bifurcation points the amplitude is proportional to Re$(\lambda)$, as in the 
unforced model. Thus the spectrum is close in form to that predicted from the 
unforced model by substituting $\langle \beta \rangle$ for the 
time-independent transmission rate \citep{bauch_earn:inter,BM10}.

There is good agreement between analytical calculations and simulations. 
Although calculations give the power spectrum as an integral, it must be 
evaluated numerically because the deterministic equations \eqref{eq:macro} 
cannot be solved in closed form; this is all carried out using the symbolic 
package Mathematica \citep{WR08}. This analysis about an annual limit cycle 
corresponds to that of \citet{BE03} except that we can derive the full power 
spectrum. They term the `resonant peak' what we describe as the deterministic 
or annual peak, and the `non-resonant peak' what we describe as the stochastic 
peaks. Their terminology is somewhat misleading, as the stochastic peak is 
generated by a resonance phenomena whereas the macroscopic peak is not.

%%%%%%%%%%%%%%%%%%%%%%%%%%%%%%%%%%%%%%%%%%%%%%%%%%%%%%%%%%%%%%%%%%%%%%%%%%%%%%
%%%%%%%%%%%%%%%%%%%%%%%%%%%%%%%%%%%%%%%%%%%%%%%%%%%%%%%%%%%%%%%%%%%%%%%%%%%%%%

\section{Period doubling and measles transitions}
\label{sec:measles}

We can use our analytic methods to help understand the dynamics and 
large-scale temporal transitions in measles epidemic patterns, first 
investigated by \cite{earn_science}. The main force in driving these 
transitions is changes in the susceptible recruitment (a mixture of changes 
in birth rates and vaccination), which can be mapped onto $R_0$. Thus a 
knowledge of the model dynamics as a function of $R_0$ can be used to explain 
the changes in epidemic patterns. Although the analysis of \cite{earn_science} 
is in good qualitative agreement with time-series data, there are a number 
of outstanding questions with regard to the interpretation of the mechanisms 
for the dynamics. We first provide a brief review of the original analysis 
and then go on to show how the stochastic dynamics of this model can be 
understood within the framework we have laid out in the previous section.  

\subsection{Review of original analysis}

\begin{figure}[t]
\centering
\includegraphics[width=\columnwidth]{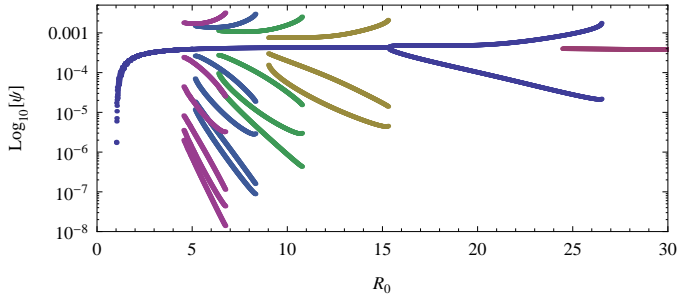}
\caption{Bifurcation diagram showing the SIR dynamics as a function of $R_0$. 
Fixed parameters: $\beta_1=0.29$, $\gamma=1/13$, $\mu=5.5\times 10^{-5}$ and 
$\eta=0$. The different period limit cycles are shown in different colours, 
which are produced by different initial conditions.}
\label{fig:earn_bif}
\end{figure}

It is acknowledged that stochasticity plays a role in the dynamics of measles, 
which can only be captured through simulation of the individual-based model. 
Fundamentally though, the analysis of these mechanisms 
by \cite{earn_science} is deterministic. Figure \ref{fig:earn_bif} shows the 
bifurcation diagram derived from the SIR equations \eqref{eq:macro}, as a 
function of $R_0$, with parameters corresponding to measles and no 
immigration ($\eta=0$). This shows the incidence sampled annually on the 1st 
of January each year, thus stable limit cycles are shown by different numbers 
of (colour coded) curves. The single curve, beginning at small $R_0$, shows 
an annual cycle which bifurcates at $R_{0}=15.5$ into two curves giving a  
biennial cycle. For values of $R_0$ lying between about 5 and 15, there are 
several sets of $n$ curves representing $n$-year cycles. 

For large $R_0$ (e.g. $R_0 \approx 30$) only an annual limit cycle exists. 
As $R_0$ is reduced a biennial limit-cycle is found; before vaccination was
introduced in England and Wales, most cities would be in this region. Higher
birth-rates might move the system back into the region with only an annual 
attractor, whereas vaccination would act to reduce $R_0$, moving it into the 
region with multiple co-existing longer period attractors. The interpretation 
put forward by Earn et al, is that stochasticity will then cause the system 
to jump between these different deterministic states \citep{Sch85b}, giving 
rise to irregular patterns. Thus, in this description, noise plays a passive 
role \citep{CRP04}.

Although peaks were seen in power spectra from simulations, which {\em appear} 
to confirm this view, there are a number of problems with this 
interpretation. The crucial aspect that is neglected is that there are no 
infectious imports included in the deterministic analysis (although presumably 
they are included in simulations). When this factor is introduced 
($\eta\ne 0$) then most of the additional structure disappears, see 
figure \ref{fig:imm_bif}a; we are left with an annual limit cycle and a 
period doubling \citep{ED94,FAG96,alonso}.

When $\eta=10^{-6}$, there is only a small region in the range $24<R_0<25$ 
where there are coexisting annual and biennial limit-cycles. As the 
immigration parameter is reduced some of the additional structure reappears; 
for example at $\eta=10^{-7}$ some of the period 3 attractors can be found
in the range $9<R_0<11$. As $\eta$ is decreased further still, more of the 
structure is found \citep{bolker_space,Nas02b}.

Immigration is an important aspect in the simulation because without it the 
disease would fade out as the minimum number of infections can go far below 
a single individual \citep{bartlett2,bolker_space,CRL+09}. In a deterministic 
analysis this term is easily omitted because the variables are continuous and 
therefore fadeout cannot happen \citep{Nas99}. This raises the question: do 
these longer period solutions have an effect on the stochastic dynamics? 
If not, how can we describe the nature of the stochastic dynamics?  We can 
use our analytic method to help clarify these questions. The power spectrum 
is especially useful as it can show up anomalous peaks from simulations. 

%%%%%%%%%%%%%%%%%%%%%%%%%%%%%%%%%%%%%%%%%%%%%%%%%%%%%%%%%%%%%%%%%%%%%%%%%%%%%%
%%%%%%%%%%%%%%%%%%%%%%%%%%%%%%%%%%%%%%%%%%%%%%%%%%%%%%%%%%%%%%%%%%%%%%%%%%%%%%

\subsection{Analytic predictions}

\begin{figure}[ht]
\centering
\includegraphics[width=\columnwidth]{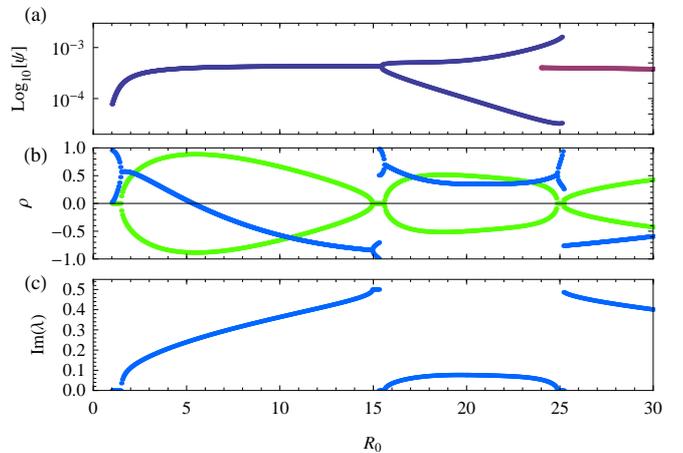}
\caption{(a) Bifurcation diagram for the SIR model with $\beta_1=0.29$ and 
$\eta=10^{-6}$. (b) The Floquet multipliers, which are in general a complex 
conjugate pair, thus we plot the real (dark blue) and imaginary (light green) 
parts separately. (c) Imaginary parts of the Floquet exponents. Note 
that in the region where there are the coexisting limit cycles ($24<R_0<25$), 
only the multipliers/exponents for the biennial cycle are shown for clarity.}
\label{fig:imm_bif}
\end{figure}

\begin{figure}[t]
\centering
\includegraphics[width=\columnwidth]{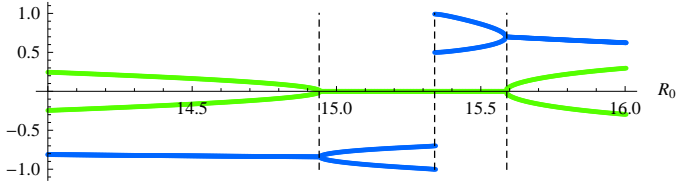}
\caption{Floquet multipliers near to the period doubling bifurcation point, 
showing the virtual-Hopf pattern. For $R_0 < 14.94$ the multipliers are a 
complex conjugate pair, with a negative real part (dark blue line); this is 
the Hopf-like region. The actual period-doubling bifurcation occurs at 
$R_0^{\text{bif}}=15.34$, where one of the multipliers becomes equal to $-1$.}
\label{fig:zoom}
\end{figure}

\begin{figure}[!ht]
\centering
\includegraphics[width=\columnwidth]{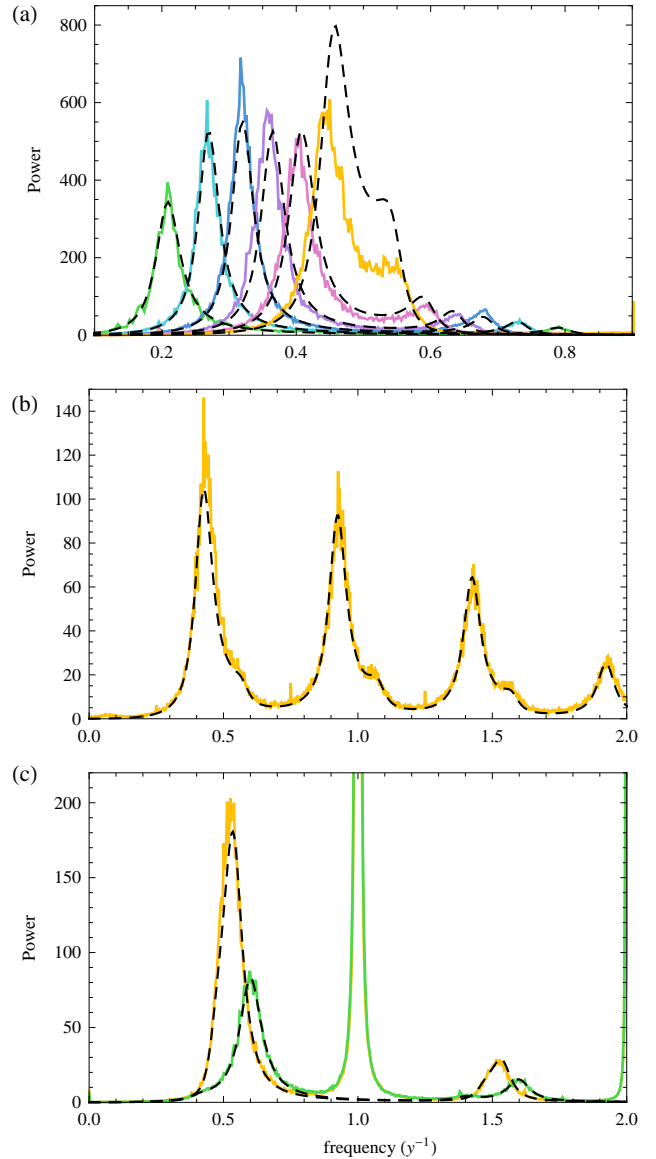}
\caption{Analytic (black dashed curves) and numerical (coloured) power spectra 
for a range of $R_0$ with $N=5\times 10^6$. In most cases the analytic and 
numerical spectra are virtually indistinguishable, apart from $R_0=14$. (a) 
Spectra before the bifurcation, $R_0=4,6,8,10,12,14$. (b) Typical biennial 
regime, $R_0=20$. Note that the stochastic peaks have been made 
clearer by subtracting the deterministic dynamics before calculating the 
power spectrum. The spectrum would otherwise be dominated by the peak at $0.5$ 
y$^{-1}$. (c) The major and minor peaks in the large $R_0$ annual regime: 
$R_0=26,30$, with the larger peaks corresponding to $R_0 = 26$ for both the 
major and minor peaks.}
\label{fig:spectra}
\end{figure}

Figure \ref{fig:imm_bif} shows the bifurcation diagram for the model presented 
in the previous section, but with $\eta=10^{-6}$, along with the Floquet 
multipliers and exponents. These parameter values will be used for the rest
of this section. Figure \ref{fig:zoom} shows the Floquet multipliers 
on a larger scale near the period doubling bifurcation point and 
figure \ref{fig:spectra} shows the analytical and numerical power spectra 
for various values of $R_0$ with $N=5\times 10^6$. Away from any bifurcation 
points there is good agreement between the analytic and the simulated spectra.

As we approach the period-doubling bifurcation point from below, the 
stochastic  oscillations follow a virtual-Hopf pattern \citep{Wis85b,Wis85}.  
This is where the oscillations first show the precursor characteristics of a 
Hopf bifurcation before changing into the precursor characteristics of a 
period-doubling. This is clearly seen in the power spectra shown in 
figure \ref{fig:spectra}. In the Hopf-like regime ($R_0 < 14.94$), the 
Floquet multipliers are a complex conjugate pair, giving rise to two peaks 
in the spectrum: a major one at frequency Im$(\lambda)$ and a minor one 
at $1-$Im$(\lambda$), as in section \ref{sec:results}. Therefore in 
figure \ref{fig:spectra}a the two peaks are most widely separated for $R_0 =4$.

As we increase $R_0$, ${\rm Im}(\lambda)$ also increases, and the major and 
minor peaks move closer together, converging at $0.5$ y$^{-1}$ when the 
multipliers become real and negative; this marks the onset of the period 
doubling regime, see figure \ref{fig:zoom}. In this regime 
$(14.94 < R_0 < 15.34$), as the multipliers are negative, so their phase is 
$\pm \pi$ and so the imaginary part of the Floquet exponents is  
$\pm \pi/2\pi T_1 = \pm 0.5$. Therefore the peak stays fixed at $0.5$ 
years$^{-1}$ as we increase $R_0$ further within this range, but the 
amplitude increases quickly. At $R_0^{\text{bif}}=15.34$ one of the multipliers
reaches $-1$ and we see a deterministic period doubling \citep{Kuz04}, and 
the size of the fluctuations grows to order $N$. Figure \ref{fig:bif} shows 
how in this way the oscillations smoothly turn into the macroscopic biennial 
limit cycle. The same pattern is seen if we hold $R_0$ fixed and increase 
$\beta_1$ to induce a period doubling. 

When the system is in the biennial regime we can still calculate the 
fluctuations about the limit cycle and get a good correspondence with analytic
predictions (figure \ref{fig:spectra}b). The positions of the peaks are now 
at $m/2 \pm \text{Im}(\lambda)$ and the spectrum changes little within this 
parameter range. The peaks at $m/2+\text{Im}(\lambda)$ are barely visible, as 
compared to the prominent peaks at $m/2-\text{Im}(\lambda)$. In the annual 
regime after the doubling ($R_0>25$), the analytic results are again very 
accurate, with stochastic peaks at frequencies $m\pm\text{Im}(\lambda)$ 
(figure \ref{fig:spectra}c). Here as well, the set of peaks at 
$m+\text{Im}(\lambda)$ are much smaller.  Note that in both of these regions 
the time-series will be dominated by the deterministic signal as the 
stochastic oscillations are much smaller than in the pre-bifurcation 
region ($R_0<15$).

%%%%%%%%%%%%%%%%%%%%%%%%%%%%%%%%%%%%%%%%%%%%%%%%%%%%%%%%%%%%%%%%%%%%%%%%%%%%%%
%%%%%%%%%%%%%%%%%%%%%%%%%%%%%%%%%%%%%%%%%%%%%%%%%%%%%%%%%%%%%%%%%%%%%%%%%%%%%%

\subsection{Near the bifurcation point}

For values of $R_0$ near the bifurcation point, the deviations between the
analytic and simulated spectra become larger (see for example 
figure \ref{fig:spectra}a; $R_0=14$). This is expected: the analysis developed 
here is essentially linear and thus predicts an unbounded increase in the 
fluctuations as we approach the bifurcation point \citep{GB05}. As the 
fluctuations become larger the linear approximation breaks down and non-linear
effects become important and act to bound the fluctuations. Going to larger 
system sizes can result in better agreement between analytic results and 
simulation, but this will always break down at some point. 

\begin{figure}[!ht]
\centering
\includegraphics[width=\columnwidth]{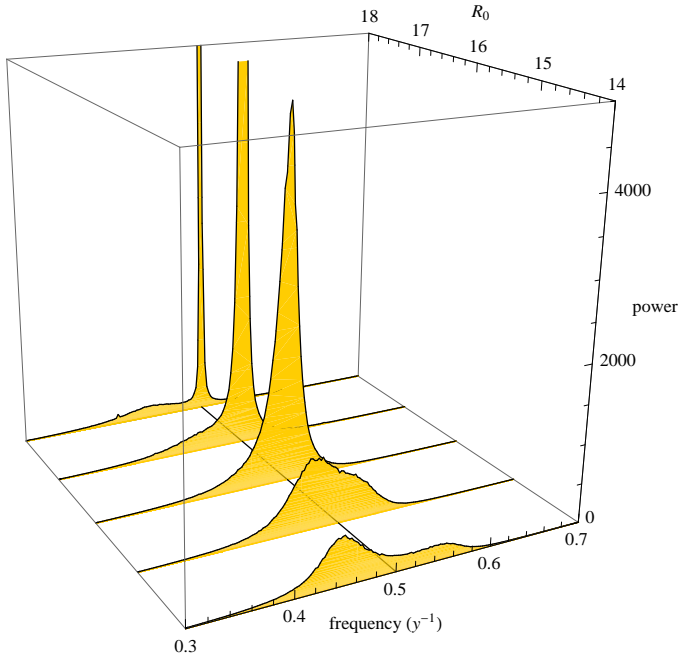}
\caption{Simulation results showing how the power spectrum of the
stochastic oscillations changes as the period doubling bifurcation point is
crossed. The peaks for $R_0=17$ and $18$ have been cropped for clarity.}
\label{fig:bif}
\end{figure}

Although the analytic approximation breaks down near the bifurcation point, 
the structure we have uncovered is still visible. Figure \ref{fig:bif} shows 
stochastic power spectra from simulations for $14<R_0<18$, as we move though 
the bifurcation point. The virtual-Hopf pattern is still clear, as predicted 
by the analysis, but the fluctuations remain bounded, growing to the same 
order as the system size \citep{van_kampen,KS03}. Within this region the 
macroscopic dynamics cannot be split into a deterministic and stochastic part 
and it is not in general possible to reconstruct the deterministic part by 
averaging over many realisations. Thus, determining exactly where the 
bifurcation takes place is difficult \citep{Wis85}. At $R_0=16$ the 
deterministic biennial peak should be observed, but is not clearly visible 
until $R_0=18$. It is possible that the bifurcation point is shifted in the 
stochastic system, but more analysis is required to determine that this is
so.

%%%%%%%%%%%%%%%%%%%%%%%%%%%%%%%%%%%%%%%%%%%%%%%%%%%%%%%%%%%%%%%%%%%%%%%%%%%%%
%%%%%%%%%%%%%%%%%%%%%%%%%%%%%%%%%%%%%%%%%%%%%%%%%%%%%%%%%%%%%%%%%%%%%%%%%%%%%

\subsection{Smaller populations}
\label{sec:small}

\begin{figure*}[!ht]
\centering
\includegraphics[]{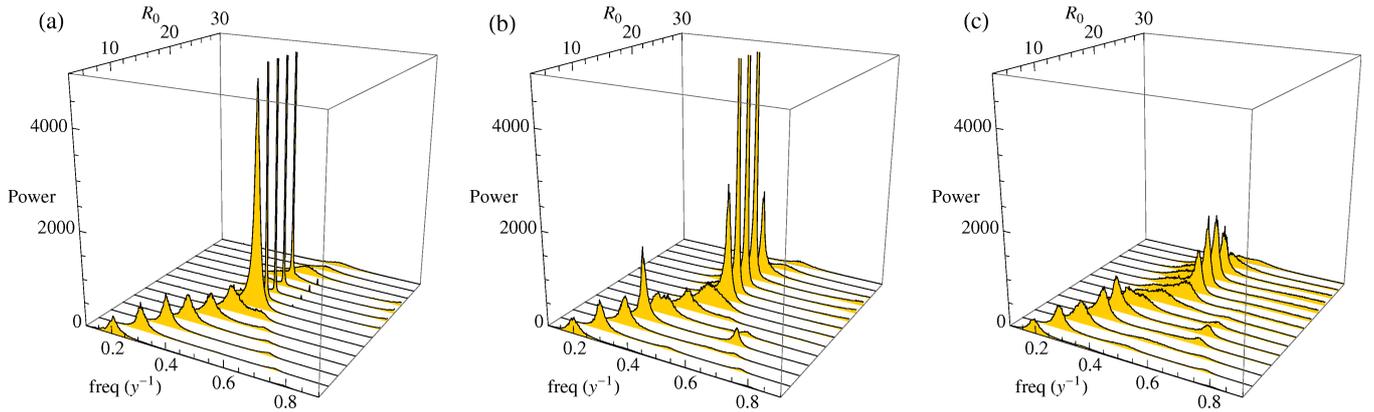}
\caption{Power spectrum through the bifurcation point for different size 
populations. (a) $N=5\times 10^6$, (b) $N=10^6$, (c) $N=5\times 10^5$. Some 
of the peaks are cropped for clarity. Notice the anomalously enhanced peak 
for $N=10^6$, $R_0=10$, see section \ref{sec:switch} for discussion of this.}
\label{fig:smallpop}
\end{figure*}

The results presented in the previous sections were for $N=5\times 10^6$, which
roughly corresponds to the largest populations we would be interested in 
modelling. Simulations of smaller populations tend to show regular deviations 
from the analytic calculations and results are sensitive to $N$, $\eta$ and 
$\beta_1$. The forcing pushes the system close to the fade-out boundary 
($I=0$), where fluctuations are non-Gaussian, and so large deviations from 
the theory are expected. Figure \ref{fig:smallpop} shows the stochastic power 
spectra from simulations, within the range $4\le R_0\le 30$ and with 
$N=5\times 10^6, 10^6$ and $5\times 10^5$.

For smaller values of $R_0$ we still clearly observe the virtual-Hopf pattern, 
but a visual inspection of the time-series shows much more irregular dynamics. 
This is due to the increased stochasticity in the smaller systems, but also 
the closeness of the fade-out boundary, where extinction and re-colonisation 
events start to have an impact on the dynamics \citep{Gri73}. This has an 
effect on the power spectra in two ways: firstly as a broadening of the power 
spectra, showing a greater range or amplified frequencies and thus a more 
irregular dynamics. Secondly the endogenous period is systematically shifted 
higher, as in unforced versions of this model \citep{STdG07}. This reflects 
the fact that the period of oscillations also depends on the re-introduction 
of the disease after fade-out \citep{bartlett2}.

The most important effect is on the fluctuations in the biennial regime after 
the period doubling. For $N=5\times 10^6$ the peaks are sharp, indicating a 
deterministic limit cycle, and the stochastic oscillations are much smaller 
(figure \ref{fig:spectra}b), hence the good predictability of these larger 
systems. For the two smaller populations this is not the case. We do not 
observe the deterministic biennial limit cycle, but instead see an enhanced 
stochastic peak and a broadening of the spectrum. The range of this 
enhanced region is also reduced. 

Although there are large deviations, having an analytical description still 
helps us interpret the dynamics at smaller $N$. Taking the average of 
Eq.~\eqref{ansatz} we obtain
\begin{equation}
\langle I \rangle=N\psi(t)+N^{\frac{1}{2}}\langle y \rangle.
\end{equation}
In the linear noise approximation, which we have used in this paper, the 
fluctuations are Gaussian and therefore $\langle y \rangle=0$. At some point 
this will break down and we must include the next order corrections, which 
will be of the order $N^{1/2}$, to the macroscopic equations. It will no 
longer be true that the mean is equal to the average \citep{van_kampen,Gri09}. 
This effect of the fluctuations on the deterministic dynamics could be enough 
to retard the onset of the biennial limit cycle and is the subject of further 
research. 

\subsection{Switching between attractors}
\label{sec:switch}

As seen from the bifurcation diagram in figure \ref{fig:imm_bif}, where 
$\eta=10^{-6}$, the only region where deterministically there are predicted 
to be two coexisting states is when $24<R_0<25$. This can be detected in 
simulations and the period of switching depends strongly on the system size. 
If the system is large it will tend to stay in the state it started in, 
because the fluctuations are not large enough compared to the mean to kick 
the system into the other state. Decreasing the system size makes this 
possible, and we see periods of annual dynamics followed by biennial and 
back to annual, where the period of switching depends on the system size. 

There is another intriguing region where we see signs of this type of 
behaviour. For $N=10^6$ and $R_0=10$ (figure \ref{fig:smallpop}b), we observe 
an enhanced stochastic peak in the spectrum with a period of 3 years. Visual 
inspection of the time-series shows regions of irregular annual oscillations 
interspersed with very regular triennial oscillations. Note that this is not 
observed in the larger or smaller systems and the power spectrum is shifted 
by the proximity to the fade-out boundary from its infinite system size limit. 
Very similar behaviour is observed for measles data from Baltimore between 
1928 and 1935 \citep{LY73,earn_science}, which has similar parameter 
values \citep{BE03}.

%%%%%%%%%%%%%%%%%%%%%%%%%%%%%%%%%%%%%%%%%%%%%%%%%%%%%%%%%%%%%%%%%%%%%%%%%%%%%%
%%%%%%%%%%%%%%%%%%%%%%%%%%%%%%%%%%%%%%%%%%%%%%%%%%%%%%%%%%%%%%%%%%%%%%%%%%%%%%

\section{Discussion}
\label{sec:discussion}

We have used an analytic approach, as well as simulations, to quantify the 
effect of stochastic amplification in the forced SIR model. The time 
dependence has been treated explicitly, instead of by using an approximation 
as in a previous paper \citep{BM10}. Because this model has a finite 
population, and therefore is inherently stochastic, it can only be studied 
`exactly' by simulation. The system-size expansion, which we use to derive 
approximate analytic solutions for this model suggests that we should view 
the population-level dynamics as being composed of a deterministic part and 
a stochastic part, where the spectrum of the stochastic fluctuations is 
intimately related to the stability of the deterministic level dynamics. 
Power spectra of these models have been known for some time, but it has not 
always been clear what the mechanisms that generate the peaks are. This is 
the main advantage of being able to calculate the power spectrum of the 
stochastic fluctuations analytically; by comparison with the simulations we 
can gain insight into the mechanisms at work.

Our analysis suggests a simple explanation for the differences seen in the 
epidemic patterns of measles and whooping cough in England and Wales both 
before and after vaccination \citep{REG99}, and which are representative of 
the two main parameter regimes for childhood diseases. The generic situation 
occurs when we are far away from the bifurcation point. Here we observe a 
deterministic annual limit cycle with stochastic oscillations, as in 
Figures \ref{fig:stoch_ex} and \ref{fig:example}. In general the form of the 
spectrum is close to that predicted by the unforced model. As already shown 
in a previous paper, this situation can account for the dynamics of whooping 
cough pre- and post-vaccination \citep{BM10}. Pre-vaccination the stochastic 
oscillations are centred on 2-3 years. Vaccination acts to shift the 
endogenous frequency lower and increases the amplitude of these fluctuations 
giving large four yearly outbreaks. 

Measles epidemics show a contrasting behaviour and represent the second 
important parameter regime, where the deterministic dynamics are near to a 
bifurcation point. Pre-vaccination, large cities such as London are in the 
regime with a deterministic biennial limit-cycle. Vaccination acts to lower 
$R_0$ and shift the system into the regime where there is an annual limit 
cycle with large stochastic oscillations. As vaccination coverage is 
increased, the endogenous period of these oscillations is also 
increased \citep{GBK01}. Measles dynamics show a strong dependence on 
population size \citep{bartlett2,GBF02}. Our analysis also offers some insight 
into this: in large populations the stochastic oscillations are very small 
compared with the deterministic biennial limit-cycle. This accounts for the 
regularity and explains why purely deterministic models capture this aspect 
so well \citep{Die76}. For smaller populations the deterministic biennial 
limit-cycle is not observed, just enhanced stochastic oscillations, thus 
accounting for the more irregular dynamics seen in these smaller populations.

Finite size effects and immigration / imports are closely related in a 
stochastic individual-based model because the population is finite. Immigration
reflects the basic fact that no population is isolated and there must be 
reintroduction of the disease if it fades out. One advantage of the 
approach which starts from an individual based model and derives the population
level model, is that the immigration terms from the stochastic model are 
automatically included in the deterministic equations on the macro-scale. As 
we have shown, this is vital, as the longer period solutions are no longer 
found when they are included, and thus removes some speculation as to the 
influence of multiple co-existing attractors. These terms are easily omitted 
in a deterministic analysis because the system size is an innocent 
parameter \citep{Nas99}. It is interesting to note the similarities between 
immigration and age-structure in these forced models \citep{Sch85}. Adding 
age-structure creates a constant pool of infectives in the infant class 
which acts to damp the dynamics \citep{BG93}, exactly analogous to how we 
model immigration. 

Owing to the importance of immigration in these epidemic systems one relevant 
outstanding question is: what form should the immigration parameter take? As 
measles dynamics can be highly synchronised it could be argued that the 
immigration parameter should reflect this \citep{bolker_space,xia_gravity,LS}. 
On the other hand for larger cities, this parameter can be viewed as an 
aggregate of many infectious encounters from varied sources and could be 
approximated as a constant. Previous work has hinted at the sort of spatial 
effects that can arise \citep{Kee00,KR02,HDF04}, but investigation of 
explicitly spatial stochastic models should be a high priority. 

The bifurcation diagrams for SEIR and SIR models are very similar, which 
justifies our use of the SIR model in this paper. The extension to uncoupled 
births and deaths would be straightforward, but would offer no further 
insight \citep{KR07}. There are technical difficulties in extending the 
method to the SEIR model because of the difference in time scales between 
the collapse onto the centre manifold \citep{SS83} and the period of forcing; 
this creates difficulties in computing the Floquet multipliers. These could 
in principle be overcome either by calculating the multipliers by a different 
method \citep{FJ91,Lus01}, or by carrying out a centre-manifold reduction
before doing the van-Kampen expansion \citep{FBS09}. The breakdown of the 
linear theory near the bifurcation point can be remedied by including 
next-to-leading order terms from the expansion of the master 
equation \citep{van_kampen}, but would result in a much more complex 
calculation. The calculations and discussions we have given here once again 
highlight the important role that simple models play in the understanding 
of complex systems. It also makes a novel contribution to the wider debate 
on the relative importance of stochastic and deterministic forces in 
ecology \citep{CRP04}. 

\section*{Acknowledgements}

\noindent
AJB thanks the EPSRC for the award of a postgraduate grant.

\appendix

\section{Expansion of the master equation}

\noindent
In this appendix we review the van Kampen system-size expansion, which
approximates the master equation (\ref{master}) by the deterministic equations 
(\ref{eq:macro}) plus stochastic fluctuations given by Eq.~(\ref{eq:langevin})
when $N$ is large. The method has been described by \cite{alonso} for the 
SIR model without forcing (i.e. where $\beta$ is independent of time), where
further details are given.

We begin by introducing step operators which allow us to express the master 
equation in a more compact form and also allow us to carry out the expansion 
in a more straightforward way. These are defined by:
\begin{equation}
\begin{split}
\mathbb{E}_I^{\pm1}\, f(S,I,t)=f(S,I\pm 1,t),\\
\mathbb{E}_S^{\pm1}\, f(S,I,t)=f(S\pm 1,I,t).
\end{split}
\end{equation}
The master equation \eqref{master} can then be written in full as,
\begin{equation}
\begin{aligned}
\frac{d}{dt}P(S,I,t)= \Biggl\{ &(\mathbb{E}_S\mathbb{E}^{-1}_I -1)\,
\Bigl[\beta\frac{S}{N}I+\eta S\Bigr] \\ 
&+(\mathbb{E}^{-1}_S -1)\,[\mu(N-S-I)]\\ 
&+(\mathbb{E}_I \mathbb{E}_S^{-1} -1)\,[\mu I] 
+(\mathbb{E}_I -1)\,[\gamma I] \Biggr \} P(S,I,t),
\end{aligned}
\label{eq:full_master}
\end{equation}
which agrees with \cite{alonso}, up to typographical errors in that paper.
The essential step in the expansion is the ansatz \eqref{ansatz}; we anticipate
that the approximate probability distribution has a mean which scales as $N$ 
and width which scales as $N^{1/2}$ \citep{van_kampen}. To expand 
Eq.~\eqref{eq:full_master} in a power series in $N^{-1/2}$, we write the step 
operators in terms of the fluctuation variables $x$ and $y$:
\begin{equation}
\begin{aligned}
\mathbb{E}_S^{\pm1} = 1 \pm N^{-\frac{1}{2}}\frac{\partial}{\partial x}+
\tfrac{1}{2}N^{-1}\frac{\partial^2}{\partial x^2}\pm \dotsb\\ 
\mathbb{E}_I^{\pm1} = 1 \pm N^{-\frac{1}{2}}\frac{\partial}{\partial y}+
\tfrac{1}{2}N^{-1}\frac{\partial^2}{\partial y^2}\pm \dotsb.
\end{aligned}
\label{eq:step_ex}
\end{equation}
Substituting these and the ansatz \eqref{ansatz} into 
Eq.~\eqref{eq:full_master} we identify a hierarchy of equations multiplied by 
different powers of $N^{-1/2}$. At leading order we find the deterministic 
equations \eqref{eq:macro} for the macroscopic variables, $\phi(t)$ and 
$\psi(t)$. At next-to-leading order we obtain a linear Fokker-Planck equation 
for the fluctuations variables $\mathbf{x}\equiv\{x,y\}$, of the form,
\begin{equation}
\frac{\partial\Pi}{\partial t} = -\sum_{i,j} K_{ij}(t)
\frac{\partial [ x_j\Pi ] }{\partial x_i} + \frac{1}{2} \sum_{i,j}
G_{ij}(t)\frac{\partial^2 \Pi}
{\partial x_i \partial x_j},
\label{FPE}
\end{equation}
where the matrices $K(t)$ and $G(t)$ depend on time through 
$\beta(t)$, $\phi(t)$ and $\psi(t)$. The Fokker-Planck equation \eqref{FPE}
is equivalent to the Langevin equations \eqref{eq:langevin} given in the main 
text \citep{van_kampen,Gar03}. Since we will not be interested in transient 
behaviour, only fluctuations about the limit cycle, the solutions to 
\eqref{eq:macro} will be those of the limit cycle, which we denote by 
$\bar{\phi}$ and $\bar{\psi}$. The explicit forms for $K(t)$ and $G(t)$ are
then given by Eqs.~(\ref{K_matrix}) and (\ref{G_matrix}) respectively. Since
$\beta(t)$, $\bar{\phi}(t)$ and $\bar{\psi}(t)$ are periodic, so are $K(t)$
and $G(t)$. 

\section{Floquet analysis}

\noindent
The analysis of the Langevin equation \eqref{eq:langevin} is not difficult
when the deterministic system approaches a fixed point at large times, and
for the SIR model this is described by \cite{alonso}. When the attractor of
the dynamics is a limit cycle, the analysis is more complicated, but can 
still be carried out to give the power spectra as integrals over known 
functions. Here we outline this analysis, following closely \cite{BGM09b},
which may be consulted for further details.

We begin by considering linear perturbations about the limit cycle, that is,
Eq.~(\ref{eq:langevin}), but without the noise which originates from the 
discreteness of the individuals. The equation describing these small 
perturbations $\mathbf{x}\equiv\{x,y\}$ is
\begin{equation}
\dot{\mathbf{x}}=K(t)\mathbf{x},
\label{eq:homo}
\end{equation}
where $K(t)$ is given by Eq.~(\ref{K_matrix}). A fundamental matrix, $X(t)$, 
is constructed from the linearly independent solutions of the homogeneous 
equation \eqref{eq:homo}, thus it satisfies the relation,
\begin{equation}
\dot{X}(t)=K(t)X(t).
\label{eq:fund}
\end{equation}
The matrix $X(t)$ is not unique and will depend on the initial conditions.
Floquet theory states that if $K(t+T_n)=K(t)$, then there exists a canonical 
fundamental matrix which can be expressed in the form $X_0(t)=P(t)Y(t)$ 
\citep{Gri90}. It has the property that the Floquet matrix, defined by
$B_{0}=X_0^{-1}(0)X_0(T_n)$, is diagonal. These diagonal elements are called 
Floquet multipliers, $\rho_i$ and play a central part in the theory. The 
matrix $P(t)$ carries the periodicity of the limit cycle, while 
$Y(t)=\text{diag}[e^{\lambda_1 t},e^{\lambda_2 t}]$, where $\lambda_i$ are 
the Floquet exponents. Using the periodicity of $P(t)$ it can be seen that 
$B = Y^{-1}(0)Y(T_n)$ and so the Floquet multipliers are related to the
Floquet exponents by $\rho_i = e^{\lambda_i T_n}, i=1,2$. Using the canonical 
form, we can derive an expression for the power spectrum in terms of the 
matrices, $P(t)$ and $Y(t)$. 

In practice, one obtains a fundamental matrix, $X(t)$ by numerically 
integrating \eqref{eq:fund} with initial condition $X(0)=\mathbb{I}$. With 
this choice of initial condition, $B=X(T_n)$. The multipliers and exponents 
are then calculated from the the eigenvalues of $B$, which allows the 
construction of the matrix $Y(t)$, since the eigenvalues are independent of 
the choice of fundamental matrix. One can then calculate the canonical form, 
$X_0(t)=X(t)S$, where the columns of $S$ are the eigenvectors of $B$. Finally 
$P(t)$ is found from $P(t)=X(0)Y^{-1}(t)=X_0(t)Y(-t)$.

Having described the basic idea behind Floquet theory, we can now return to
the Langevin equation \eqref{eq:langevin}, which is an inhomogeneous linear 
equation with periodic coefficients. We can use Floquet theory to construct a 
solution to this by adding a particular solution to the general solution of 
the corresponding homogeneous equation \eqref{eq:homo} \citep{Gri90}. This 
gives,
\begin{equation}
\mathbf{x}(t)=X(t)X^{-1}(t_0)\mathbf{x}_0 + 
X(t)\int_{t_0}^t X^{-1}(s)\mathbf{f}(s) ds,
\label{eq:sol1}
\end{equation}
with initial condition $\mathbf{x}(t_0)=\mathbf{x}_0$. We are interested in 
the steady state solutions, when transients have damped down, thus we can 
ignore the first part of Eq. \eqref{eq:sol1} and set the initial time to the 
infinite past, $t_0\rightarrow-\infty$. Taking the case where $X(t)$ is
$X_0(t)=P(t)Y(t)$, one finds using the properties of the diagonal matrix 
$Y(t)$, that
\begin{equation}
\mathbf{x}(t)=P(t)\int_{-\infty}^t Y(t-s)P(s)^{-1}\mathbf{f}(s) ds.
\label{eq:sol2}
\end{equation}
The correlation matrix is defined as
$C(t+\tau,t)=\langle\mathbf{x}(t+\tau)\mathbf{x}^{T}(t)\rangle$, which using
Eq.~(\ref{eq:sol2}) may be written as
\begin{eqnarray}
C(t+\tau,t)&=&P(t+\tau)\int_{-\infty}^{t+\tau}\int_{-\infty}^t 
Y(t+\tau-s)P(s)^{-1}G(s) \nonumber \\
&\times& \delta(s-s')(P^{-1}(s'))^T Y(t-s')^T\,ds'\,ds\,P(t)^{T}, \nonumber \\
\end{eqnarray}
where $\langle\mathbf{f}(s)\mathbf{f}^{T}(s')\rangle=G(s)\delta(s-s')$. 
Integrating over the delta function, the result will depend on the sign of 
$\tau$. If we take $\tau\ge0$ then the integration region is $-\infty<s<t$, 
giving
\begin{equation}
\begin{aligned}
C(t+\tau,t)= P(t+\tau)\int_{-\infty}^t Y(t+\tau-s) 
\Gamma(s) Y(t-s)^T \,ds \, P(t)^T,
\end{aligned}
\end{equation}
where we have defined
\begin{equation}
\Gamma(s)=P(s)^{-1}G(s)(P^{-1}(s))^T,
\end{equation}
which will have the periodicity of the limit cycle. Next we make a change of 
variables, $s\rightarrow t-s'$, which gives
\begin{equation}
\begin{aligned}
C(t+\tau,t)=  P(t+\tau)\int_0^{\infty} Y(\tau+s') 
\Gamma(t-s') Y(s')^T \,ds' \, P(t)^T.
\end{aligned}
\end{equation}
The form of $Y$ means we may write $Y(\tau + s')=Y(\tau)Y(s')$, and so the 
integral that we need to evaluate is given by
\begin{equation}
\Phi(t) \equiv \int_0^\infty Y(s)\Gamma(t-s)Y^T(s)\, ds.
\end{equation}
Using the periodicity of the matrix $\Gamma(t-s)$, this integral can be recast 
as a finite one over the period of the limit cycle:
\begin{equation}
\Phi_{ij}=\frac{1}{1-\rho_i\rho_j}\int_0^{T_n} \Gamma_{ij}(t-s) 
e^{(\lambda_i+\lambda_j)s} ds.
\label{appB_result1}
\end{equation}
Therefore, the final expression for the correlation matrix is
\begin{equation}
C(t+\tau,t)= P(t+\tau)Y(\tau)\Phi(t)P(t)^T.
\label{appB_result2}
\end{equation}
So we can obtain the correlation matrix as an integral, but this has to be 
evaluated numerically because the neither the limit-cycle solutions nor $P(t)$
can be obtained in closed form.

\bibliographystyle{model2-names}
\bibliography{refs.bib} % remove before submission

\end{document}